\documentclass[aps,twocolumn]{revtex4}
\usepackage[dvips]{graphics,graphicx}
\begin{document}
\title{Atomic current in optical lattices: Esaki-Tsu equation revisited}
\author{Andrey R. Kolovsky}
\affiliation{Kirensky Institute of Physics and
Siberian Federal University, 660036 Krasnoyarsk, Russia}
\affiliation{Department of Physics, University of Crete and Foundation for
Research and Technology-Hellas, 71003 Heraklion, Greece}
\date{\today}

\begin{abstract}
The paper discusses the master equation approach to derivation of the Esaki-Tsu equation for drift current. It is shown that the relaxation term in the master equation can be identified by measuring the velocity distribution of the carriers. We also show that the standard form of the relaxation term, used earlier to derive Esaki-Tsu equation, predicts unphysical velocity distribution and suggest a more elaborated relaxation term, which is argued to correctly capture the effect of bosonic bath in experiments on atomic current in optical lattices.
\end{abstract}
\maketitle

{\em 1.} Recently much attention has been payed to the transport of cold atoms in optical lattices \cite{Ott04,Scot04,Stof04,70,Gust07,Snoe07,69,Brud07}. This is not only due to the fact that this system mimics electrons in a crystal lattice, thus allowing studies a variety of transport phenomena of the solid state physics. More importantly, the system of cold atoms in optical lattice offers a possibility for direct measuring of the quantities, which were earlier measured only indirectly, and opens an access to the parameter regimes, which were inaccessible with solid crystals. This addresses a number of questions, which may require a revision of the results obtained earlier in the field of the condensed matter physics. In this brief report we discuss the Esaki-Tsu equation for the drift current, suggested by Esaki and Tsu in 1970 with respect to the biased semiconductor super-lattices \cite{Esak70}. Recent experimental \cite{Ott04} and theoretical \cite{69,Brud07} studies of the atomic dynamics in tilted optical lattices indicate that this equation also holds, at least qualitatively, for the cold atoms. The aim of the present work is to explore the `degree of uncertainty' in the Esaki-Tsu equation which, as it will be shown below, comes from different models for interactions between the system (the carriers, in what follow) and its environment (the bath). We shall also show that in the case of cold atoms one can get information about the actual interactions by measuring the velocity distribution of the carriers instead of measuring the net current.

\bigskip
{\em 2.} First we recall the reader the quantum-mechanical derivation of the Esaki-Tsu equation \cite{remark0}. The starting point is the master or Liouville equation on the reduced density matrix of the carriers,
\begin{equation}
\label{1}
\frac{d \hat{\rho}}{dt}=-i[\widehat{H}_0,\hat{\rho}]
+{\cal L}(\hat{\rho}) \;,
\end{equation}
where $\widehat{H}$ is the carriers single-particle Hamiltonian in the tight-binding approximation,
\begin{equation}
\label{2}
\widehat{H}=\widehat{H}_0 + F\sum_l l |l\rangle\langle l| \;,\quad
\widehat{H}_0=\frac{J}{2}\sum_l \left( |l+1\rangle\langle l| +h.c.\right)
\end{equation}
and ${\cal L}(\hat{\rho})$ the relaxation term, which takes into account the effect of a bath. (To simplify equations we set the lattice period and Planck's constant to unity.) It is further assumed that that for $F=0$ the bath brings system into the thermal state $\bar{\rho}_0$ and that this process is characterized by a single relaxation constant $\gamma$, i.e.,
\begin{equation}
\label{3}
{\cal L}(\hat{\rho})=-\gamma(\hat{\rho} - \bar{\rho}_0) \;,\quad
\bar{\rho}_0 \sim \exp(-\beta\widehat{H}_0) \;.
\end{equation}
Clearly the thermal density matrix $\bar{\rho}_0$ is diagonal in the quasi-momentum basis $|k\rangle=L^{-1/2}\sum_l \exp(i2\pi kl/L) |l\rangle$. On the other hand, the Hamiltonian $\widehat{H}$ is diagonal in the basis of the Wannier-Stark states $|m\rangle=\sum_l {\cal J}_{l-m}(J/F) |l\rangle$ (here ${\cal J}_n(z)$ are Bessel functions of the first kind), with eigenvalues forming the Wannier-Stark ladder. Using the latter basis the stationary solution of the master equation (\ref{1}) with the relaxation term (\ref{3}) reads \cite{Mino04},
\begin{equation}
\label{4}
\bar{\rho}=\frac{\bar{\rho}^{(0)}_{m,m'}}{1+i(m'-m)(F/\gamma)} \;, \quad
\bar{\rho}^{(0)}_{m,m'}=\frac{1}{L}\frac{{\cal I}_{m-m'}(\beta J)}{{\cal I}_0(\beta J)} 
\end{equation}
where ${\cal I}_n(z)$ are the second kind Bessel functions. Finally, substituting (\ref{4}) into expression for the mean carrier velocity,
\begin{displaymath}
\bar{v}={\rm Tr}[\hat{v}\bar{\rho}] \;,\quad 
\hat{v}=\frac{J}{2i}\sum_l \left(|l+1\rangle\langle l| - h.c.\right) \;,
\end{displaymath}
one recovers the Esaki-Tsu equation with the temperature dependence given in the prefactor $f(\beta)$,
\begin{equation}
\label{5}
\frac{\bar{v}}{v_0}= f(\beta)\frac{F/\gamma}{1+(F/\gamma)^2} \;,\quad
f(\beta)=\frac{{\cal I}_1(\beta J)}{{\cal I}_0(\beta J)} \;.
\end{equation}

At this point we note that the result (\ref{5}) heavily relies on a particular choice of the relaxation term (\ref{3}), which by no means can be considered as well justified. For this reason we consider a more general relaxation term,
\begin{equation}
\label{6}
{\cal L}(\hat{\rho})=-\frac{\gamma}{2}\sum_{s,q} W(s,q)
\left[\hat{\rho}\hat{\sigma}^{\dagger (s,q)}\hat{\sigma}^{(s,q)}\right. 
\end{equation}
\begin{displaymath}
\left.- 2\hat{\sigma}^{(s,q)}\hat{\rho}\hat{\sigma}^{\dagger (s,q)} 
+ \hat{\sigma}^{\dagger (s,q)}\hat{\sigma}^{(s,q)} \hat{\rho}\right] \;,
\end{displaymath}
which is explicitly written here in the Lindblad form \cite{Lind76}. In Eq.~(\ref{6}) the operators $\hat{\sigma}^{(s,q)}$ in the quasimomentum representation are given by the matrices
\begin{equation}
\label{7}
\sigma^{(s,q)}_{k,p}=\delta_{k,s}\delta_{p,q}
\end{equation}
and the coefficients $W(s,q)$ have sense of the transition rates between different quasimomentum states in the absence of static forcing. The only fundamental restriction on the coefficients $W(s,q)$ is 
\begin{equation}
\label{7a}
W(q,s)=e^{\beta(E_q-E_s)}W(s,q) \;,\quad 
E_q=-J\cos\left(\frac{2\pi q}{L}\right) \;,
\end{equation}
which insures relaxation of an arbitrary initial state to the thermal equilibrium. In all other aspects the entries $W(s,q)$ are model specific. In the rest of the paper we analyze two of these models, which are especially useful for understanding physics behind Eq.~(\ref{5}).

\bigskip
{\em 3.} It is convenient to rewrite the master equation (\ref{1}) with the relaxation term (\ref{6}) as an equation on matrix elements of the density matrix in the quasimomentum representation,
\begin{equation}
\label{8}
\dot{\rho}_{k,p}=-[\widehat{H},\hat{\rho}]_{k,p} 
-\frac{\gamma}{2}\left(\sum_s W(s,k)+\sum_s W(s,p)\right)\rho_{k,p} \;,
\end{equation}
\begin{equation}
\label{9}
\dot{\rho}_{k,k}= -[\widehat{H},\hat{\rho}]_{k,k} 
- \gamma\left(\sum_s W(s,k) \rho_{k,k} + \sum_q W(k,q) \rho_{q,q}\right) 
\end{equation}
First we consider the case where the transition rates are given by $W(s,q)=\delta_{s,0}$. Then, assuming for a moment $F=0$, the balance equation (\ref{9}) simplifies to 
\begin{equation}
\label{10}
\dot{\rho}_{0,0}=\gamma\sum_{k\ne 0}\rho_{k,k} \;,\quad
\dot{\rho}_{k,k}=-\gamma\rho_{k,k} \;.
\end{equation}
Physically this corresponds to zero temperature of the bath and selection rules, where any quasimomentum state relax directly into the ground state. We would like to stress that the considered choice of the coefficient $W(s,q)$ is the only case where the relaxation term (\ref{6}) coincides with (\ref{3}). For all other choices a reduction of (\ref{6}) to the simple form (\ref{3}) requires some approximations or is impossible in principle. 

Let now $F\ne0$. Assuming the limit of an infinite lattice, where $\rho_{k,k}(t)=\rho(k,t)$ is a continuous function of the quasimomentum, the coherent evolution term modifies Eq.~(\ref{10}) as
\begin{equation}
\label{11}
\frac{\partial \rho(k,t)}{\partial t}=F\frac{\partial \rho(k,t)}{\partial k}
-\gamma\left(\rho(k,t)+\delta(0) \int_0^{2\pi}\rho(k,t) dk\right) \;.
\end{equation}
Since $\int \rho(k,t) dk=1$ due to the normalization, the stationary solution of (\ref{11}) is given by
\begin{equation}
\label{12}
\bar{\rho}(k)=\frac{2\pi\gamma}{F}\frac{1}{1-\exp(2\pi\gamma/F)}
\exp\left(\frac{\gamma}{F}k\right) \;,
\end{equation}
with discontinuous jump at $k=0$. (Note that $\rho(k,t)$ is a periodic function of the quasimomentum.) Substituting this stationary solution into expression for the mean velocity, $\bar{v}=J\sum_k \sin(2\pi k/L)\bar{\rho}_{k,k}$, we recover Eq.~(\ref{5}) for the net current.

\begin{figure}
\center
\includegraphics[width=8cm, clip]{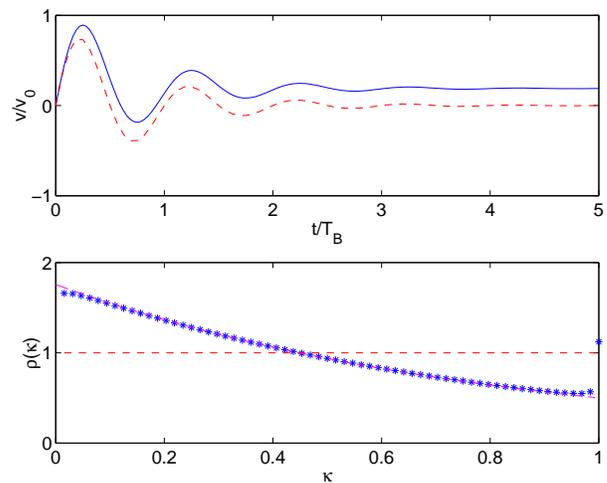}
\caption{Numerical analysis of the model No.1. The mean velocity as the function of time (solid line in the upper panel) and stationary distribution over the quasimomentum (asterisk in the lower panel), compared with the analytical result (dash-dotted line). Parameters are $F=0.2$, $\gamma=0.04$, $k_BT=0$, and $L=64$ (periodic boundary conditions). For the sake of comparison dashed lines show the mean velocity and stationary distribution at infinite temperature.} 
\label{fig1}
\end{figure}

\begin{figure}
\center
\includegraphics[width=8cm, clip]{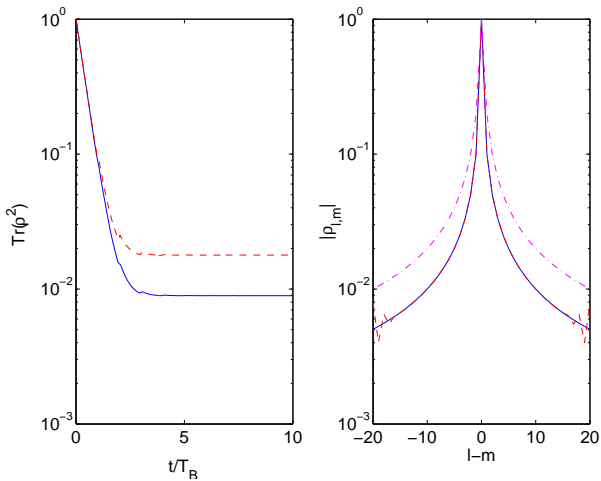}
\caption{Decay of coherence (left) and normalized matrix elements of the stationary density matrix across the main diagonal (right) for $L=64$ (dashed line) and $L=128$ (solid line). The other parameters are the same as in Fig.~1.} 
\label{fig2}
\end{figure}

As an illustration to the above analysis and to test the code Fig.~\ref{fig1} and Fig.~\ref{fig2} show numerical solution of the master equation for $W(s,q)=\delta_{0,s}$. In our numerical approach we solve the equation in the Wannier basis for a finite lattice with periodic boundary conditions. Note that for $F\ne0$ imposing the periodic boundary conditions requires the gauge transformation, which can be also seen as interaction representation with respect to the Stark term in the Hamiltonian (\ref{2}). We have checked that this procedure does not introduce artifacts and all numerical results converge with an increase of the lattice size $L$.
 
The solid line in the upper panel in Fig.~\ref{fig1} shows time evolution of the mean carrier velocity for the initial conditions given by the ground state of the system (the Bloch wave with zero quasimomentum). It is seen that the static force induces Bloch oscillations, which decay to a finite $\bar{v}$ after a transient time $\sim 1/\gamma$. Then, by plotting $\bar{v}$ as the function of $F$, we reproduce the Esaki-Tsu dependence  (results are not shown). The lower panel in Fig.~\ref{fig1} depicts quasimomentum distribution at the end of numerical simulations, -- a nice correspondence with the analytical solution (\ref{12}) is noticed. We also would like to mention the crucial effect of forcing on the system coherence, which we characterize by the linear entropy $c(t)={\rm Tr}[\hat{\rho}^2]$. Without static force the system relax into the ground state and $c(t)$ approaches unity. Opposite to this, if static field is on, the density matrix tends to a diagonal matrix in the momentum representation and almost diagonal matrix in the coordinate representation with the diagonal elements $\rho_{l,l}\approx1/L$. Thus $c(t)$ approaches $1/L$, as it is illustrated in the left panel in Fig.~\ref{fig2}. Nevertheless, it would be wrong to say that static force makes the system completely incoherent. A weak coherence inherited from coherent Bloch oscillations is preserved in the form of small but nonzero off-diagonal elements $\bar{\rho}_{l,m}$ with characteristic profile shown on the right panel of Fig.~\ref{fig2}. This profile is compared with the simplest estimate
\begin{displaymath}
\bar{\rho}_{l,m}=\frac{1}{L}\frac{1}{1+i(F/\gamma)(l-m)} \;,
\end{displaymath}
which one obtains from (\ref{1}) and (\ref{3}) by setting the hopping matrix element in the Hamiltonian (\ref{2}) to zero.

\bigskip
{\em 4.} The main drawback of the model considered above is that it predicts discontinuous distribution function for the carrier velocity (which also holds for a finite temperature). Next we analyze a more realistic model, which is free from this drawback. It is instructive to begin with the case of infinite temperature \cite{remark1}, where the master equation can be obtained from the first principles. For example, if the bath consists of Bose atoms in an optical lattice deep enough to justify the Bose-Hubbard model and carriers (for example, Fermi atoms, as in Ref.~\cite{Ott04}) interact with Bose atoms according to $\widehat{H}_{int}=\epsilon\sum_l |l\rangle\langle l| \hat{n}_l$, the relaxation term in the master equation (\ref{1}) has the form \cite{69}, 
\begin{equation}
\label{13}
{\cal L}(\hat{\rho})_{l,m}=\gamma(1-\delta_{l,m})\rho_{l,m} 
\end{equation}
in the Wannier basis, or
\begin{equation}
\label{14}
{\cal L}(\hat{\rho})_{k,p}=
-\gamma\rho_{k,p}+\frac{\gamma}{L}\sum_q \rho_{k+q,p+q} 
\end{equation}
in the Bloch basis, where $\gamma\sim \epsilon^2{\bar n}^2/\hbar J_B$ with $\bar{n}$ being the density of Bose atoms, $J_B$ hopping matrix elements for Bose atoms, and $\epsilon$ the interaction constant for collision-like interactions of Bose atoms with the carriers \cite{remark2}. It is easy to see that the relaxation operator (\ref{14}) corresponds to the choice $W(s,q)=1/L$ in the Lindblad form (\ref{6}). As a solution of the master equation with relaxation term (\ref{14}) one gets the decaying Bloch oscillations and uniform stationary distribution for the carrier velocity.
\begin{figure}
\center
\includegraphics[width=8cm, clip]{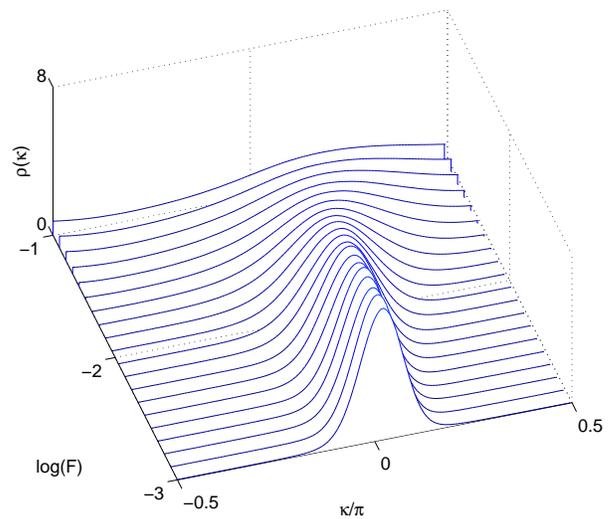}
\caption{Numerical analysis of the model No.2. Stationary distributions $\bar{\rho}(k)$ for different values of $F$. Parameters are $\gamma=0.08$, $J/k_BT=10$, and $L=128$ (periodic boundary conditions).} 
\label{fig3}
\end{figure}
\begin{figure}
\center
\includegraphics[width=8cm, clip]{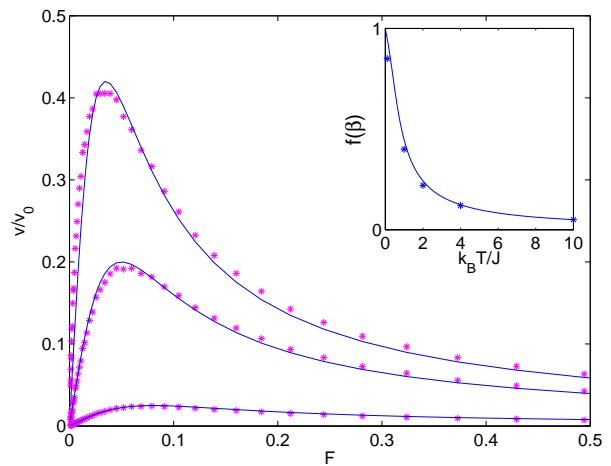}
\caption{Net current for $J/k_BT=10$, $J/k_BT=1$, and $J/k_BT=0.1$ (asterisk), fitted by the Esaki-Tsu equation (solid lines). The other parameters are the same as in Fig.~3. The inset shows maximally possible current, as compared with the temperature dependence $f(\beta)$ in Eq.~(\ref{5}).} 
\label{fig4}
\end{figure}

Since $W(s,q)=1/L$ implies equal exchange rates for any pair of the quasimomentum states, a reasonable model for a finite temperature is the set of $W(s,q)$ where transitions from any lower to any upper energy level are suppressed by the Boltzmann factor according to Eq.~(\ref{7a}). Figure \ref{fig3} shows the stationary distributions $\bar{\rho}(k)$ calculated on the basis of this model. Unlike in the previously considered case $W(s,q)=\delta_{s,0}$, the stationary distributions are now smooth functions of $k$, with the center of gravity shifted to positive quasimomenta ($F>0$). In addition to Fig.~\ref{fig3}, Fig.~\ref{fig4} depicts the mean carrier velocity as the function of the static force. It is seen that obtained dependencies are well approximated by the Esaki-Tsu formula (\ref{5}), although they do not coincide with (\ref{5}) exactly. In particular, with temperature decrease the position of the maximum is found to move towards smaller $F$, while in the standard model it is always at $F=g$. One can also see small deviations from the standard model in the value of maximal current, shown in inset of Fig.~\ref{fig4}.

\bigskip
{\em 5.} In conclusion, we have analyzed the Esaki-Tsu equation for the drift current with respect to ongoing experiments on atomic current in the tilted or accelerated optical lattices. It is shown that different models of carrier interactions with the bath result in essentially the same (Esaki-Tsu) dependence of the net current on the static force magnitude and the bath temperature. This proves one more time the universal character of the Esaki-Tsu equation. However, different models predict drastically different distribution functions for the carriers velocity. Since velocity distribution can be directly measured in the laboratory experiments with cold atoms, this additional information can be used for identifying the details of carrier-bath interactions.

We also mention that the presented analysis questions the common believe that the drift current is obligatory a non-Markovian process \cite{69,Mino04,Brud07}, where the system keep memory about its initial state. Indeed, the above discussed carrier dynamics is explicitly Markovian, nevertheless, it results in the Esaki-Tsu dependence for the drift current.

The author gratefully acknowledge discussions with A.~Buchleitner and A.~V.~Ponomarev and financial support within the 6-th Framework Programme (project FP6-032980-2).


\newpage


\begin{thebibliography}{99}
\bibitem{remark0}
We note that originally this equation was obtained by using the semiclassical approach. The presented quantum derivation due to \cite{Mino04} closely follows the original semiclassical argumentation.
\end{thebibliography}
\end{document}